\documentclass[preprint,superscriptaddress]{revtex4-1}
\usepackage{calligra}
\DeclareMathAlphabet{\mathcalligra}{T1}{calligra}{m}{n}

\usepackage{amsmath} 
\usepackage{amssymb} 
\usepackage{amsfonts}

\usepackage{graphicx} 

\usepackage{array}
\usepackage{multirow}
\usepackage{color}
\usepackage{transparent}
\usepackage{float}
\usepackage[mathscr]{eucal}

\usepackage{natbib}

\renewcommand{\bm}[1]{{\bf #1}}
\newcommand{\kB}{k_\mathrm{B}}
\newcommand{\lB}{l_\mathrm{B}}
\newcommand{\ie}{{\it{i.e.}}}

\def\Eq{Eq.}

\def\Fig{Fig.}

\begin{document}

\title{Potential of mean force and transient states in polyelectrolyte complexation} 

\author{Xiao Xu}
\affiliation{Institut f{\"u}r Physik, Humboldt-Universit{\"a}t zu Berlin, Newtonstr.~15, 12489 Berlin, Germany}
\affiliation{Institut f\"ur Weiche Materie und Funktionale Materialien, Helmholtz-Zentrum Berlin, Hahn-Meitner-Platz 1, 14109 Berlin, Germany}

\author{Matej Kandu\v{c}}
\affiliation{Institut f\"ur Weiche Materie und Funktionale Materialien, Helmholtz-Zentrum Berlin, Hahn-Meitner-Platz 1, 14109 Berlin, Germany}

\author{Jianzhong Wu}
\affiliation{Department of Chemical and Environmental Engineering, University of California, Riverside, California 92521, United States}

\author{Joachim Dzubiella}
\affiliation{Institut f{\"u}r Physik, Humboldt-Universit{\"a}t zu Berlin, Newtonstr.~15, 12489 Berlin, Germany}
\affiliation{Institut f\"ur Weiche Materie und Funktionale Materialien, Helmholtz-Zentrum Berlin, Hahn-Meitner-Platz 1, 14109 Berlin, Germany}
\email{joachim.dzubiella@helmholtz-berlin.de}

\begin{abstract}
The association between polyelectrolytes (PEs) of the same size but opposite charge is systematically studied in terms of the potential of mean force (PMF) along their center-of-mass reaction coordinate via coarse-grained, implicit-solvent, explicit-salt computer simulations. 
The focus is set on the onset and the intermediate, transient stages of complexation. 
At conditions above the counterion-condensation threshold,  the PE association process exhibits a distinct sliding-rod-like behavior where the polymer chains approach each other by first stretching out at a critical distance close to their contour length, then 'shaking hand' and sliding along each other in a parallel 
fashion, before eventually folding into a neutral complex. The essential part of the PMF for highly charged PEs can be very well described by a simple theory based on sliding charged `Debye--H\"uckel' rods with renormalized charges in addition to an explicit entropy contribution owing to the release of condensed counterions.  Interestingly, at the onset of complex formation, the mean force between the PE chains is found to be discontinuous, reflecting a bimodal structural behavior that arises from the coexistence of interconnected-rod and isolated-coil states. These two microstates of the PE complex are balanced by subtle counterion release effects and separated by a free-energy barrier due to unfavorable stretching entropy.  

\end{abstract}

\maketitle

\section{Introduction}

Polyelectrolyte complexes (PECs) consisting of oppositely charged polymer chains have  been studied for decades owing to their fundamental importance in biophysics and technological applications~\cite{Dubin,JaspervanderGucht2011}. Common examples include delivery vehicles for gene therapy and oral vaccination \cite{Kissel, Lankalapalli2009,KrishnenduRoy1999,JihanZhou2013}, neurofilament association~\cite{beck}, membrane support for filtration processes~\cite{M.B.Dainiak1998,VladimirA.Izumrudov1998}, functional coatings~\cite{laschewsky}, and highly-ordered macromolecular structures for water treatment and mining~\cite{JaeyoungLee2004,MartinSwanson-Vethamuthua1997,AndersLarsson1999}. A major focus of previous investigations has been on the physicochemical nature of the complex-forming polyelectrolytes (PEs) and environmental conditions that control the structural and topological properties.  The asymmetry between polyions of opposite charges leads to PECs with a broad range of conformations and size distributions, suggesting that the PEC size and topology~\cite{Dubin,JaspervanderGucht2011} as well as assembly kinetics~\cite{Cheng1998, SaskiaLindhoud2012} can be controlled  by alternating the polymer chemistry as well as solution conditions. 

In particular, polycations for gene delivery are designed to associate with nucleic acids so that the gene transfer vectors are able to overcome the intracellular barriers such as the plasma membrane, the endosome, and the nuclear membrane~\cite{Kissel}. Successful gene delivery depends upon the ability of the vector to adopt different metastable structures or possess certain properties at different intracellular environments. It has been shown, for instance, that the complex of small interfering RNA~(siRNA) with poly(amidoamine) (PAMAM) undergoes aggregation and precipitation at different stages of delivery, and the dynamic behavior is reported to possibly facilitate the transaction efficiency~\cite{JihanZhou2013}. Different binding characteristics were reported when a DNA helix interacts with synthetic polymers of different charge density (e.g., polyethylenimine (PEI) and poly-L-lysine (PLL))~\cite{Ziebarth2009}. Because the effective polymer charge density is closely linked to counterion-condensation effects~\cite{Ray1994,Manning1998,MarkusDeserno2000,Qian2000,Naji2006}, the appearance of different transient structures depends on the ionic strength and counterion valence. In that respect, Stevens~\cite{Stevens2001} observed toroids and rods formed from a bead-spring PE chain in the presence of polyvalent counterions that could have major impacts on the kinetics of PE complexation.

In order to understand complex formation between PEs on a detailed molecular level (in both pair or many-body PE systems), monomer-resolved computer simulations are a powerful tool. Examples include atomistically-resolved systems~\cite{Hoda2009, Ziebarth:2009, Qiao:2010, Elder2011, Antila1,Antila2}  as well as coarse-grained models~\cite{Muthu:1994, Winkler:2002, yoshikatsuhayashi2004, Ou2006, R.S.Dias2011, Peng2015}.  In addition to detailed structural and topological insights into the final complexed state, those studies indicate that the primary driving forces of PE complexation for weakly-charged PEs arises as expected from electrostatic attraction.  By contrast, association between highly charged PEs is dominated by the release of condensed counterions if the polymer charge density is above  the threshold value for the phenomenon of counterion-condensation to occur~\cite{Ray1994,Manning1998,MarkusDeserno2000,Qian2000,Naji2006}.  As systematically shown by Ou and Muthukumar using coarse-grained simulations and a mean-field lattice model, counterion-release effects have remarkable consequences on the thermodynamics of complexation, which is of mostly entropic origin for highly charged PEs, while mostly enthalpic for the weakly charged ones \cite{Ou2006}. The entropic effect on the binding free energy appears to be consistent with the classical Record-Lohman picture for PE pair complexation \cite{M.ThomasRecordJr1976,DavidP.Mascotti1990}. 

Despite the extensive theoretical simulation work on PEC formation on a pair level, to the best of our knowledge, little attention has been given to the variation of the system's free energy and structures along the PE association pathway.  The previous simulation studies indicate very fast and cooperative association processes once the isolated PE coils come into first  contact~\cite{Muthu:1994, Winkler:2002,Ou2006}. Here, apparently, transient but very distinctly stretched states of the PEs appear that govern the beginning of the kinetic association towards the final complex.  Very recently, Peng and Muthukumar~\cite{Peng2015}  calculated the potential of mean force (PMF)   along the center-of-mass  distance reaction coordinate and reported a constant force of attraction (linear PMF) for a large distance range comparable to the PE contour length. A systematic exploration of the PMFs for different PE charge densities, however, is still absent, 
as well as their quantitative description by a tractable theory.  Also the nature of the transient states right at the onset of attraction and during
complexation is not fully characterized, despite the need for a better understanding of PEC metastable states and 
preceding association kinetics, e.g., in gene delivery and therapy~\cite{Kissel, Ziebarth:2009, JihanZhou2013}.

The present work is concerned with the process of complex formation between two oppositely charged PEs right at the onset of 
the association and the intermediate range before collapsing into the final state.  For this, we investigate a pair of two oppositely charged 
PEs of the same size and charge density using implicit-water, explicit-salt Langevin simulations of a generic (bead-spring) PE 
model for various charge densities.  Our main focus is placed on the PE configurations and the association free energy as well as 
the number of released ions along the center-of-mass distance reaction coordinate. Above the counterion-release threshold charge
density we find very distinct sliding-rod association pathways accompanied by an essentially constant mean force. We show that
the latter can be described by a combination of a simple Debye-H\"uckel-like theory for associating rods with the counterion-release
entropy that dominates for PE chains with large charge densities. Importantly, we observe that the mean force is discontinuous at the onset of complexation. 
We explain this observation by a bimodal structural behavior that arises from the coexistence of interconnected-rod and isolated-coil states. 
The two intermediate states of the PE complex are balanced by subtle counterion-release mechanisms and separated by a free-energy 
barrier due to unfavorable stretching entropy of the PE chains. Possible consequences of our findings are briefly discussed in the final section of this work.

\section{Models and theoretical background}

\subsection{Polyelectrolyte model and simulations}

Consider an aqueous solution containing two polyelectrolyte (PE) chains of the same size but opposite charge in the presence of salt ions at a finite concentration. We treat the solvent implicitly via a uniform dielectric background with a permittivity constant of water at room temperature, $\epsilon_r = 78.44$. The PE chains are represented by a coarse-grained beads-on-the-string model \cite{Rubinstein1987}. Approximately, each bead represents a monomer for a realistic PE chain. In addition to electrostatic interactions, the polymer beads interact with each other in terms of the Lennard-Jones (LJ) potential 
\begin{equation}
U_\textrm{LJ}(r) = 4\epsilon_{\rm LJ}\left[\left(\frac{\sigma_{\rm LJ}}{r}\right)^{12}-\left(\frac{\sigma_{\rm LJ}}{r}\right)^6\right]
\label{bp}
\end{equation}
with a diameter $\sigma_\textrm{LJ} = 0.3$\,nm and an energy $\epsilon_\textrm{LJ}  = 0.1~\kB T$. Monovalent salt ions are modeled explicitly as charged beads with the same LJ potential as that for the PE beads. The valency of small ions is $z_\pm =\pm 1$.

The PE chain connectivity is imposed by a harmonic potential 
\begin{equation}
U_\textrm{bond}(l) = k_\textrm b(l-l_0)^2,
\label{bp}
\end{equation}
where $l$ represents the distance between consecutive beads, and $l_0=0.4$\,nm is the equilibrium bond length. The spring 
constant is $k_\textrm b = 4100$\,kJ mol$^{-1}$\,nm$^{-2}$. To account for the polymer backbone flexibility, we restrain the bond angle also by a harmonic potential
\begin{equation}
U_\textrm{angle} = k_\textrm a(\theta - \theta_0)^2,
\label{ap}
\end{equation}
where $\theta$ is the angle determined by a triplet of nearest-neighboring beads, and $\theta_0 = 120^\circ$ is its prescribed equilibrium value. The potential constant is $k_\textrm a = 418$ kJ mol$^{-1}$ rad$^{-2}$~\cite{CemilYigit2015}. 

In this work we consider relatively short PE chains with the total number of beads $N_\textrm {b} = 25$, close to the degree of polymerization for PEs used in some related experimental studies~\cite{ShunYu2015,J.-L.Popot2003,FrankWiesbrock2005,E.Pezron1989,JihanZhou2013}.
The contour length of the PE chains is $L_c \simeq (N_\textrm b - 1) l_0\sin(\theta_0/2) \simeq 8.3$\,nm, where the sine function takes care of the bond angular constraints in a moderately stretching regime. Each bead carries a bare Coulomb charge $|q_\textrm b|=e |z_b^\pm|$, with $e$ being the unit charge and $|z_b^+|=|z_b^-|\equiv z_b$ the valency. The total charge of each PE chain is $|Q_\textrm b|$ = $N_\textrm{b}|q_\textrm b|$. Since we consider complex formation between two anti-symmetric PE chains, the final complex is electroneutral. The two polymers have the same properties except with opposite charges, that is, $Q_\textrm b > 0$ for one PE chain, and $Q_\textrm b < 0$ for the other.  

Our computer simulations are based on the Langevin dynamics (LD). The equation of motion for each bead is give by~\cite{Zwanzig1954}
\begin{equation}
m_i\frac{d^2\bm{r}_i}{dt^2} = -m_i\xi_i\frac{d\bm{r}_i}{dt} + \bm\nabla_{i}U + \bm{R}_i(t),
\label{Langevin}
\end{equation}
where $m_i$ and $\xi_i$ are the mass and the friction coefficient of the $i$th bead (or small ion), respectively.  All beads have the same unit mass, which is set to minimize the inertia effects and enhance sampling. The choice of particle mass does not affect the equilibrium properties of polymers studied in this work. The total potential energy of the entire system is the sum of all the contributions $U = U_\textrm{Coul}+U_\textrm{LJ}+U_\textrm{bond} + U_\textrm{angle} $. The random force ${\bm{R}}_i(t)$ in eq.~(\ref{Langevin}) has zero mean and its autocorrelation function satisfies the fluctuation-dissipation theorem~\cite{Zwanzig1965}
\begin{equation}
 \langle \bm{R}_i(t) \cdot \bm{R}_j(t') \rangle = 2m_i\xi_i\kB T\delta(t-t')\delta_{ij}.
\end{equation}
The Langevin friction is chosen as $\xi_i = 1.0$ ps$^{-1}$ such that it dissipates energy at constant temperature $T = 300$~K on the time scale much faster than those governing the dynamics of the polymer system. To integrate the equations of motion, we employ the leap-frog algorithm with a time step of $2$~fs. The simulations are carried out by the {GROMACS 4.5.4} software package~\cite{BerkHess2008}.

In the production runs, we use a cubic box with a side length of $L = 30$\,nm with periodic boundary conditions in all three directions.
The center of mass translation of the system is removed in every $10$th integration (time) step. The electrostatic interactions are treated with the Particle-Mesh-Ewald method (PME)~\cite{UlrichEssmann1995}, where the long-range potential is evaluated in the reciprocal space using the Fast Fourier Transform (FFT) with a grid spacing $0.12$\,nm and the cubic interpolation of the fourth order. We use a cutoff radius $r_\textrm{cut} = 3.0$\,nm for the short-range electrostatics and the LJ interactions. The choice of the cutoff is verified by reference simulations with increased cutoff value $r_\textrm{cut}=5.0$\,nm.  For PE chains with the highest charge density  simulated in this work (Manning parameter $\xi = 2.31$, defined below), where the electrostatic interactions are most significant, the potential of mean force~(PMF) curves deviated less than $6\%$ from the results treated with $r_\textrm{cut} = 3.0$\,nm. In all simulations, the PEs are immersed in a salt solution with $N_i = 325$ pairs of cations and anions, resulting into a salt bulk concentration of $c_0 = N_i/L^3 \simeq 20$ mM.

\subsection{Charge densities and counterion condensation}

According to the Onsager--Manning--Oosawa theory~\cite{Ray1994,Manning1998,MarkusDeserno2000,Qian2000,Naji2006},  the net charge of a highly-charged PE is renormalized by condensation of the surrounding counterions.  The condensed and free ions should be considered separately as two distinctive states. While the counterions condensed at the surface of the PE backbone form a strongly-correlated liquid,  the  free ions in the diffusive double layer \cite{Grahame1947,I.Prigogine2007} can be approximately described by a Debye--H\"uckel~(DH)-like mean-field theory. The analytic approach will be utilized later in this work to develop a simple expression for the PMF between two anti-symmetric PEs.  

The propensity of counterion condensation is characterized by the Manning parameter~\cite{Ray1994,Manning1998,MarkusDeserno2000,Qian2000,Naji2006}
\begin{eqnarray}
\xi = |z_\pm| \lB(\lambda/e).
\end{eqnarray}
Here, $|z_\pm|$ stands for the counterion valency (\ie, $|z_\pm|=1$ in our case), $\lambda =|q_\textrm b|/[l_0\sin( \pi/3)]$ is the {\it bare} line charge density, and  $\lB = e^2/(4\pi \epsilon_0\epsilon_r \kB T)$ stands for the Bjerrum length, which is $\lB = 0.71$~nm for water at room temperature as used in our simulations.
Note that due to thermal fluctuations of the flexible chain, the bare charge density is defined as above strictly only in the low temperature limit.

The Onsager--Manning--Oosawa theory~\cite{Ray1994,Manning1998,MarkusDeserno2000,Qian2000,Naji2006} predicts that
counterion condensation occurs if $\xi > 1$, i.e., if the Manning parameter exceeds unity.  In this work, we consider the Manning parameter in the range from $\xi = 0.29$ to $\xi = 2.31$, corresponding to the bare charge density from $\lambda = 0.412$ to $3.3$~e/nm or the monomeric charge between $q_{\textrm b} = 0.14286e$ and $q_{\textrm b} = 1.14286e$, respectively. When $\xi > 1$, the two-state model predicts a fraction $\theta=1-1/\xi$ of counterions in the condensed state. The conventional model for ion distribution in the diffusive double layer is based on a cylindrical cell with the PE chain placed at the center \cite{IrinaA.Shkel2001,Denton2010}. If the polymer backbone is treated as a cylinder, the effective, renormalized charge density of such a cylinder with $\xi>1$ is then
\begin{eqnarray}
\lambda_r = e/l_{\rm B}, 
\label{xiq}
\end{eqnarray}
resulting in a Manning parameter of unity.  The cell model provides simple expressions for the condensation threshold radius $r_0$, the Manning parameter $\xi$, and the fraction $\theta$ of condensed counterions~\cite{Heyda2012}. It predicts that the fraction $\theta$ of condensed ions is independent of the cell size~\cite{Manning1969,MarkusDeserno2000,Qian2000,Naji2006}. When $\xi<1$, counterion condensation does not occur and the PE bare charge and the effective charge are the same.

In order to estimate the number of condensed counterions, $N$, in our simulations, we count an ion as condensed if it is located within a radial distance $r_0$ from any bead of a PE chain with the opposite electrostatic charge. To avoid double counting, each ion is counted only once according to the nearest distance to the polymer beads. Understandably, the radius $r_0$ lacks a rigorous definition (see, e.g., the discussion in Refs.~\cite{MarkusDeserno2000,Liu:2002}), in particular for flexible PEs in an electrolyte solution.  In this work, we propose an unambiguous procedure to account for counterion condensation: the radius $r_0$ is fixed such that the Manning prediction of $N = \lambda/e (1-1/\xi)L_c$ is obeyed for each PE chain. This procedure yields $r_0 = 0.5, 0.56, 0.63, 0.68$, and $0.74$\,nm for $\xi =1.15, 1.44, 1.73, 2.02$, and $2.31$, respectively. These values are reasonable because it is known that $r_0$ increases monotonically with increasing $\xi$ and that it should be identical to the effective PE (modeled as rod) radius for $\xi=1$~\cite{MarkusDeserno2000}, which is about $0.3-0.4$~nm for our PE model. Similar values of $r_0$ can be estimated from an inspection of the radial distribution of counterions (not shown) around the PE beads (see, e.g., Refs.~\cite{Liu:2002, ShunYu2015, CemilYigit2015}).

\subsection{PMF calculation}

\begin{figure}[h]
\includegraphics[scale=0.45]{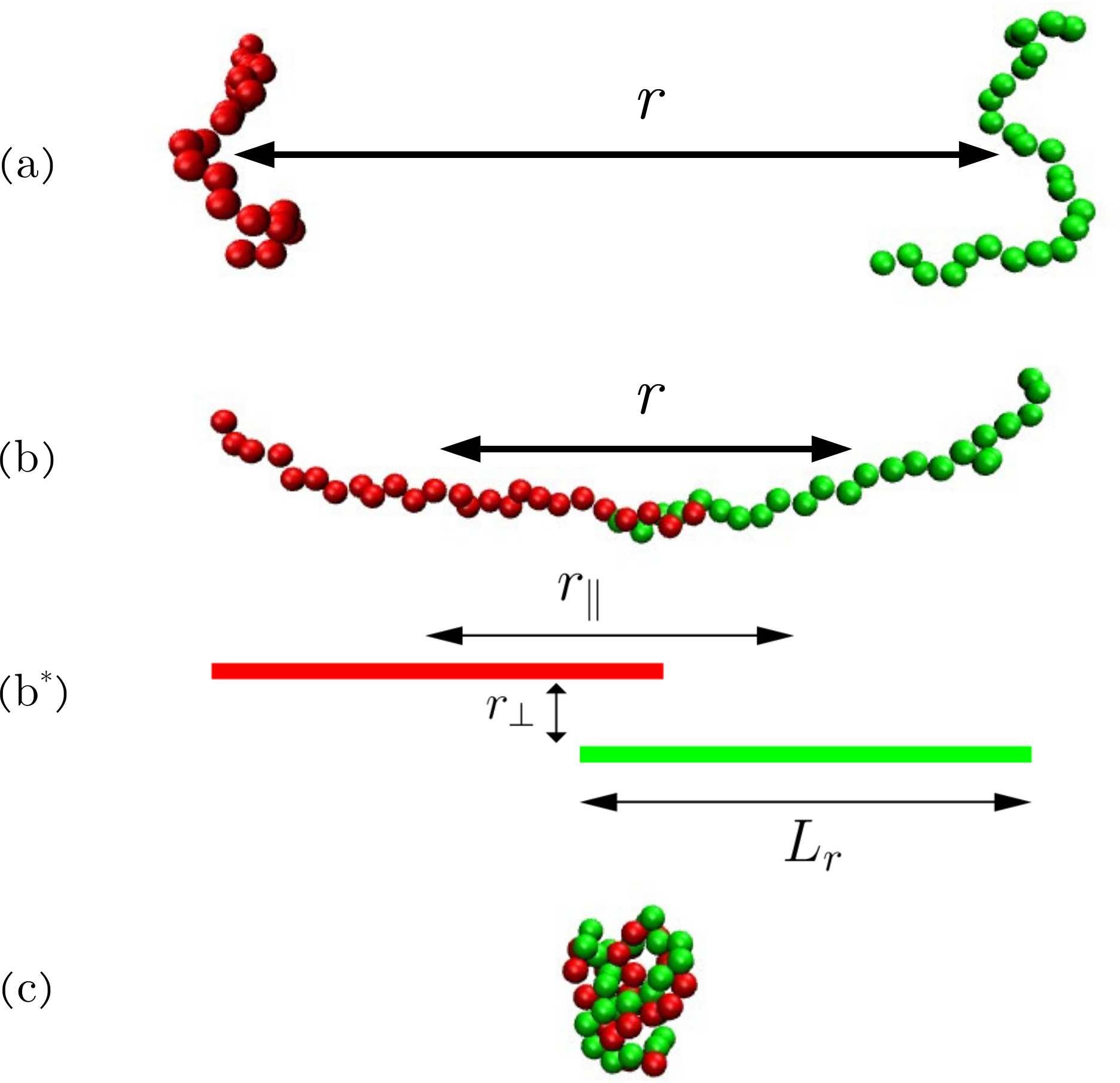}
\caption{Snapshots of the PE binding process, depicting the situation of (a) initially separated PEs in a coil-like state, (b) 
the `handshake' at the onset of complexation between elongated PE chains, and (b$^*$) a schematics thereof where the whole situation 
is projected on parallel, sliding-rods. The red rod depicts the polyanionic chain, whereas the green rod the polycationic chain in 
our DH rod model (defined below). Panel (c) depicts the final, entangled PE complex. The reaction coordinate $r$ corresponds to the COM distance between two PEs of opposite charge. Each PE chain consists of 25 monomeric units each of the same absolute partial charge. The 
Manning parameter in this example is $\xi = 2.02$, and the PE chains are immersed in a salt solution of the  bulk concentration 
$c_0 \simeq 20$ mM (ions not shown).}
\label{snapshot}
\end{figure}

To obtain the PMF between two oppositely-charged PEs, we use steered Langevin Dynamics (SLD) simulations with the 'pull-code' as implemented in {GROMACS}~\cite{BerkHess2008}.  Here, the distance $r$ between the centers-of-mass (COM) of the two PE chains serves as the reaction coordinate, which is constrained by an external harmonic potential. The fictitious potential is time-dependent such that it exerts a pulling force to steer the PE chains moving in a prescribed direction with a prescribed velocity $v_p$. An exemplifying snapshot of two PEs before association is shown in Fig.~\ref{snapshot}(a). 

We have empirically tested a variety of steering velocities to make sure that the PE drift is slow enough to sample the equilibrium state, i.e., the simulation results are independent of the choice of $v_p$.  For all results reported in this work, the steering velocity $v_p = 0.2$\,nm ns$^{-1}$ is used along with the harmonic force constant $K = 2500$ kJ mol$^{-1}$\,nm$^{-2}$. During the simulation, we steer the constrained PE chains approaching each other from a well separated state ($r\sim 12$~nm) to the the final state ($r \sim 0.5$~nm). The friction force $f_f = -m\xi_i v_p$ is subtracted from the constraining force and the result is averaged within a 
specific interval of the discrete spacing $\Delta r$ to obtain the mean force of the interaction potential. After that the PMF profile is acquired with a backward integration. Since one of the PE chains is radially constrained in a three-dimensional space, we need to subtract the center-of-mass translational entropy ~\cite{Neumann1980,BerkHess2006,Kalcher2009,MarkusDeserno2000}
\begin{equation}
W(r) = W^{I}(r) + (D -1)\kB T \ln r,
\end{equation}
where $W^{I}(r)$ is the integrated mean force and $D=3$ is the dimensionality of the external constraint. 

We use an analogous procedure to calculate the free energy of stretching a single PE chain. In that case, the distance between the head and tail monomers from the same PE chain serves as the reaction coordinate.

\subsection{PE alignment order parameter}

In order to characterize and demonstrate the mutual alignment of two highly-charged PEs during complex formation, we introduce the order parameter $m_{pp}$ defined as
\begin{eqnarray}
m_{pp}^i = \left\langle\frac{|{\vec{t}_{ic}} \cdot {\vec{t}_{cc}}|}{||\vec{t}_{ic}||\,||\vec{t}_{cc}||}\right\rangle.
\label{mpp}
\end{eqnarray}
Here, $\vec{t}_{ic}$ is the intra-PE direction vector connecting one terminal bead and the central bead of the same polymer $i$, with $i=1,2$, whereas $\vec{t}_{cc}$ is the inter-PE direction vector linking the central beads of both PEs. The sign $||\ldots||$ represents the norm of a vector, and $\langle \ldots \rangle$ denotes the ensemble average. For stretched, rod-like polymer configurations, we expect parallel alignment of both vectors $\vec{t}_{ic}$ with $\vec{t}_{cc}$, \ie,  the PE chains are aligned with themselves and with the connection axis. In that case, the order parameter approaches unity for both PE chains, $m_{pp}^i \lesssim 1$.  If two parallel rods approach each other in a perpendicular direction, or if the rods are perpendicularly approaching, or if the orientations of the PEs are uncorrelated,  we would expect an average value of the order parameters much closer to $m_{pp}^i \simeq 1/2$.  We will plot and discuss the distance-resolved value for $m_{pp}(r) = [m_{pp}^1(r)+m_{pp}^2(r)]/2$, which is averaged over the two PE chains. 

Moreover, we monitor the distance-resolved end-to-end distance $R_{ee}(r)$ of the PEs as they approach each other. A stretching of the PE chains can then be identified simply from the increase of $R_{ee}(r)$ at a certain COM distance. 

\subsection{An analytical model for the PMF between highly-charged PE chains}

To attain a better understanding of the thermodynamic driving forces and the mechanisms of PE--PE association, we now introduce a simple electrostatic model for the PMF between highly-charged  PE chains ($\xi>1$). 
We assume that the PEs adopt an isolated, coil-like state when they are far apart, as sketched in~Fig.~\ref{snapshot}(a). In this case, the PEs interact roughly with a DH-like potential $V_\textrm {DH}(r) \propto Q_r^2\l_{\textrm B}\exp(-\kappa r)/e^2$, where $Q_r \simeq 11.7e$ is the renormalized net charge of each PE and $\kappa = \sqrt{8\pi l_{\textrm B} c_0}$ the usual 
DH screening parameter. For distances larger than the contour length $r\gtrsim L_c$ this accounts only for a few $k_{\textrm B} T$ of attraction and will be discarded in the following discussion. We further assume that during their association at smaller distances $r \lesssim L_c$, the two PEs are in a parallel sliding conformation, as depicted in ~Fig.~\ref{snapshot}(b$^*$), until they reach a final state with $r$ close to zero. In the latter, the PE chains have collapsed into a globular complex, cf. Fig.~\ref{snapshot}(c). Consequently, we assume that the PMF is dominated by a parallel sliding process for a wide range of separations. 

We model the two approaching PE chains in their fully stretched association configuration with the simplest analytically tractable model, that is, two parallel, infinitely thin, and oppositely-charged rods of length $L_r$. The latter corresponds to the effective length of a stretched PE in the longitudinal direction that should be close to the contour length $L_c$ but its precise value will be fitted to the simulation results for the PMFs. The COM distance of the two rods is again denoted by $r$, with the parallel component $r_\parallel$ and the perpendicular component $r_\perp$, as shown in~Fig.~\ref{snapshot}(b$^*$).  The latter is kept fixed and reflects the closest distance of the two PEs, which is related to the LJ diameter $\sigma_\textrm{LJ}$. We now assume that the major interaction contribution to the total free energy arises only from the neighboring parallel segments of length $L_r-r$ from each rod.  In the following, we will evaluate in detail the electrostatic contribution based on a DH approach and additionally account for the purely entropic contribution of the counterion release effect in the scenarios with $\xi>1$.

We first evaluate the DH energy corresponding to the electrostatic interaction between two parallel, partially neighboring rods. Assuming $r_\perp\ll L_r$, we may neglect the edge effects and approximate the pair potential between the neighboring segments as those from infinitely long rods. With the pairwise additive approximation, we can derive the overall interaction energy then as~\cite{Manning1969}
\begin{equation}
\beta W_\textrm{DH}(r)= (r-L_r )\lB (1-\theta)^2 (\lambda/e)^2  \int^{\infty}_{-\infty} \frac{e^{-\kappa \sqrt{x^2+r^2_\perp}}}{\sqrt{x^2+r^2_\perp}}  dx.
\label{first}
\end{equation}
Here, $\beta=1/ (\kB T)$, $\lambda$ is the bare line charge of a PE rod, which can be partially neutralized by condensed counterions. The fraction of neutralization is $\theta=0$ for $\xi<1$ and $\theta = 1 - 1/\xi$ if $\xi>1$. Note that in the latter case, the effective charge density stays as $e/\lB$, 
irrespective of the COM distance~$r$. Performing the above integral, yields~\cite{Manning1969, Poon2006}
\begin{equation}
\beta W_\textrm{DH}(r)=
\frac{2}{\lB}(r - L_r) K_0(\kappa r_\perp) \times
\left\{
\begin{array}{c l}      
   \xi^2& \quad\textrm{for} \,\, \xi < 1,  \\
   1& \quad\textrm{for} \,\, \xi \ge 1,  \\
\end{array}\right.
\label{nr}
\end{equation}
where $K_0(\kappa r_\perp)$ is the Bessel function of the second kind. According to \Eq~(\ref{nr}), the electrostatic DH part of the PMF
is linear in the center-of-mass distance $r$, which is a logical outcome of our asummption, since only neighboring segments of the parallel aligned rods
contribute. Note again that because of the charge renormalization effect for $\xi\geq 1$,
this contribution does not explicitly depend on the charge density.

For $\xi \geq 1$, in addition to the direct electrostatic interactions between polyions, we may estimate the contribution to the free energy of PE complexation due to the release of counterions. For association between two oppositely-charged PEs with the Manning parameter $\xi \gg 1$, the free energy is dominated by the entropically favored release of condensed counterions. The idea was first proposed by Lohman and coworkers~\cite{M.ThomasRecordJr1976,DavidP.Mascotti1990} and supported later by other theoretical investigations including coarse-grained 
(but explicit-salt) computer simulations~\cite{Ou2006, CemilYigit2015}. The entropy gain upon releasing $n$ bound counterions is
\begin{eqnarray}
\Delta S_\textrm{ion} =  n\kB\,\ln(c_\textrm b/c_0), 
\end{eqnarray}
where $c_\textrm b$ corresponds to the effective density of the condensed ions in the vicinity of the PE and $c_0$ is the ion bulk density.
The corresponding free energy gain due to counterion release is then given by
\begin{eqnarray}
\beta W_\textrm{ion} =  -n\,\ln(c_\textrm b/c_0).
\label{freee_cr}
\end{eqnarray}

For complex formation between two PEs of opposite electric charge, each PE chain carries condensed counterions of a line density $\lambda\theta$, where $\theta$ is the fraction of the PE charge neutralization by counterions. When the PE segments approach each other, all counterions in the overlapping region are liberated and released into the bulk solution. The number of released ions is $n=2\times \lambda \theta (L_r - r)/e$. In terms of the Manning parameter, $n$ can be expressed as
\begin{equation}
n(r) 
\simeq 
\left\{
\begin{array}{c l}      
   0&\quad\textrm{for} \,\, \xi <1, \\
   \frac 2\lB (\xi-1)(L_r - r)&\quad\textrm{for} \,\, \xi \ge 1.  \\
\end{array}\right.
\label{nnr}
\end{equation}

The density of the bound ions~\cite{CemilYigit2015,ShunYu2015} is defined by the number of condensed counterions $N$ and their occupied volume $V$, i.e., $c_\textrm b=N/V$.
The condensed counterions reside around each rod at a radial distance between the inner radius $\sigma_\textrm{LJ}$ and the outer radius $r_0$. The occupied volume thus corresponds to that of two hollow cylinders, $V=2s_c(L_r - r)$, where $s_c=\pi(r_0^2-\sigma_{LJ}^2)$. Accordingly, the density of the bound ions for $\xi>1$ can be estimated as
\begin{equation}
c_\textrm b = \frac{(\xi-1)}{\lB s_c}.
\label{densityratio}
\end{equation}
The free energy contribution from the released counterions is then given by
\begin{eqnarray}
\beta W_\textrm{ion} =  2 \frac{(\xi -1)}{\lB}(L_r - r)\ln\Bigl(\frac{\xi-1}{\lB s_c c_0 }\Bigr).
\label{freee_cr}
\end{eqnarray}  
The above expression is valid only for $\xi>1$; otherwise, $W_\textrm{ion} = 0$ since there are no counterions condensed on the PE surface. For a salt concentration $c_0 = 20$~mM and, for instance, $\xi$ between $1.15$ and $2.31$, Eq.~(\ref{densityratio}) predicts $c_b\simeq$ 0.7~M and 2.12~M, respectively. That corresponds to 3.6$-$4.7 $\kB T$ of dissociation free energy per single released ion, respectively.

Finally, we sum up the DH contribution $W_\textrm {DH}$ and the counterion release $W_\textrm{ion}$, viz.
\begin{equation}
\beta W_{\textrm {theo}}(r)=\left\{
\begin{array}{c l}      
   \cfrac{2}{\lB}\, \xi^2  K_0(\kappa r_\perp) (r - L_r) &\\\textrm{ for }  \xi \le 1,\\
   \cfrac{2}{\lB} \left[ K_0(\kappa r_\perp) + (\xi -1)\ln(\frac{\xi-1}{\lB s_c c_0})\right](r - L_r)  &\\ \textrm{ for } \xi > 1.  \\
\end{array}\right.
\label{fr_total}
\end{equation}
Equation.~(\ref{fr_total}) predicts a PMF between PE chains to be linearly dependent on the COM distance $r$. As previously concluded in related simulations of PE complexation~\cite{Ou2006}, the PMF is dominated by electrostatic enthalpy for $\xi<1$ and counterion-release entropy for $\xi \gg 1$. 
It should be noted that Eq.~(\ref{fr_total}) is valid only for a finite length of overlapping rod segments, $r<L_r$. It neglects smaller contributions for larger distances $r>L_r$. 
The separation distance between the rods $r_{\perp}$ is approximately equal to the LJ diameter $\sigma_\textrm{LJ}$. However, $r_{\perp}$ could be smaller than that, since a bead of one PE chain can sit in the region between the two neighboring beads of the other PE chain. In this work, we obtain the value for $r_{\perp}$ by the best fit of \Eq~(\ref{fr_total}) to the simulation data for $\xi = 1.0$, yielding $r_{\perp} = 0.28$~nm.  We use this value for all other $\xi$-values. 

\section{Results}

\subsection{Structure of the PEs during association}

\begin{figure}[h]
\includegraphics[scale=0.7]{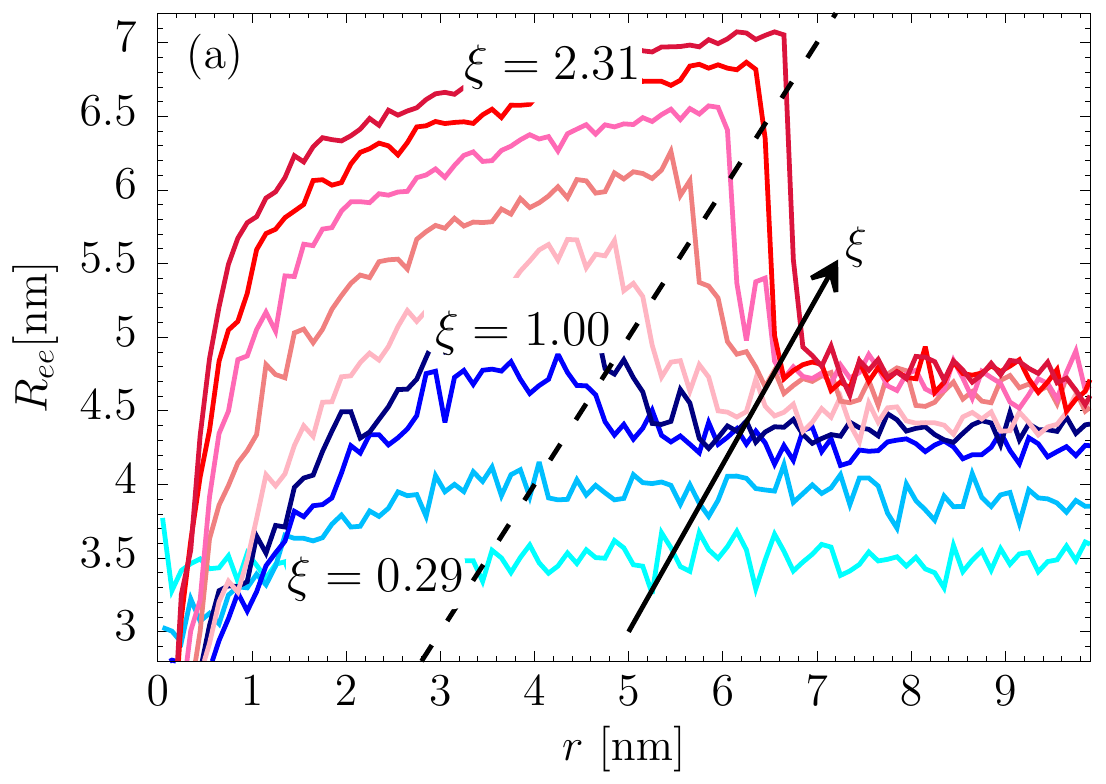}
\includegraphics[scale=0.7]{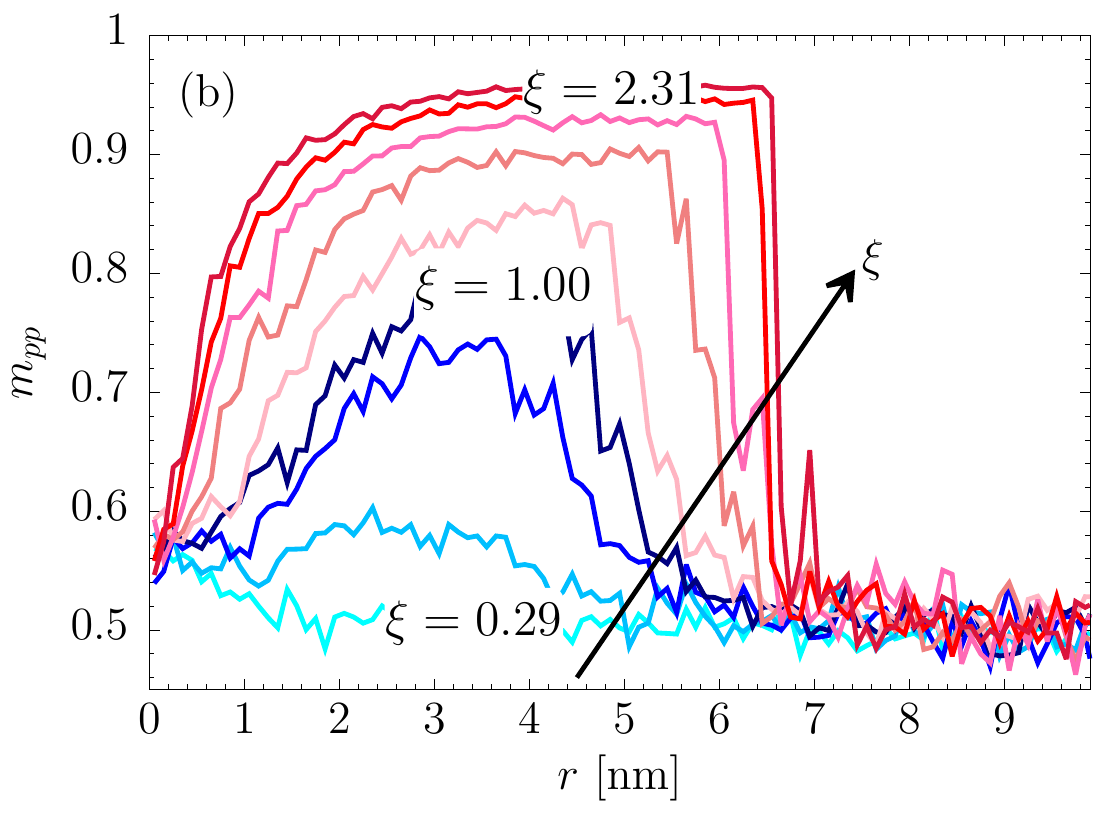}

\caption{(a) The PE end-to-end distance $R_{ee}$ as a function of the PE--PE COM distance $r$ for various values of the Manning parameter ranging from $\xi = 0.29$ to $\xi = 2.31$ at salt concentration $c_0=20$ mM. For $\xi>1$, $R_{ee}$ jumps up at a certain distance $r^*$ to the value $R^*_{ee}$. The black dashed diagonal line  shows the function $R_{ee}(r) = r$, revealing a linear correlation $R^*_{ee} \approx r^*$. (b) The PE--PE orientation order parameter $m_{pp}$, \Eq~({\ref{mpp}}), is monitored as
a function of the COM distance $r$ for various values of the Manning parameter at salt concentration $c_0=20$ mM. 
For both panels, $\xi = 0.29, 0.58, 0.87, 1.00, 1.15, 1.44, 1.73, 2.02$ and $2.31$ (from bottom to top). The arrows signify the trend for increasing $\xi$.
}
\label{re}
\end{figure}

A first insight into the configurations of the PE chains along their approach to association can be attained by analyzing the distance-resolved end-to-end distance $R_{ee}(r)$  of the individual PEs. Figure~\ref{re}(a) shows the average $R_{ee}(r)$ of the antisymmetric PEs with different Manning parameters. At large separations, where the PE chains are in a coil-like state and independent of each other, cf.~\Fig~\ref{re}(a), $R_{ee}(r)$ is constant. The effective size of the coil increases with increasing $\xi$ due to self-electrostatic repulsion within the PE chain. For highly charged chains, $\xi \gg 1$, a significant and steep, almost discontinuous increase of the end-to-end distance can be observed at COM distances close  to the contour length $r^{*} \simeq 5-7$~nm. The exact value of the critical distance $r^*$ depends on $\xi$. 
 For larger $\xi$, the critical distance increases towards the maximum contour length and the magnitude as well as steepness of the jump grow. This behavior strongly indicates stretching of the PE chains towards each other and 'handshake'~\cite{beck} at a critical distance $r^*$ to assume a maximum end-to-end length $R_{ee}^*$ as shown in the snapshots in \Fig~\ref{snapshot}(b). An observed linear relation $R_{ee}^*\propto r^*$, cf. \Fig~\ref{re}(a), further consolidates that proposition. At small distances, $r\lesssim 2$~nm, the PE chains seem to collapse again to with the $R_{ee}(r)$ approaching the values that correspond to the isolated coil states.

The behavior of $R_{ee}(r)$ for the associated states strongly correlates with the alignment order parameter $m_{pp}$, shown in \Fig~\ref{re}(b). At large distances $r \gtrsim 6.5$\,nm, the PE chains are independent of each other and uncorrelated in any alignment, as indicated by \Fig~\ref{snapshot}(a). In that case, the order parameter tends to the value $m_{pp}\simeq 1/2$. At the intermediate distances $r\simeq 5-7$\,nm, again cf.~\Fig~\ref{snapshot}(b), the ends of different PEs jump together and the polymer chains become elongated and $m_{pp}$ rises to high values, indicating parallel alignment of stretched PEs. The order parameter in this regime increases with increasing $\xi$. Particularly for $\xi = 2.31$, the order parameter reaches $m_{pp}=0.95$, implying that the two PEs get almost completely stretched and perfectly aligned. At smaller distances, $r\lesssim 2$~nm, the order parameter tends to the value $m_{pp}\simeq1/2$, and the previously extended PEs collapse into a compact globule. In a relatively large spatial regime roughly appearing at distances between $1-2$\,nm $< r < 5-7$\,nm (with the exact values depending on $\xi$), both PEs slide along each other in a stretched and parallel configuration, with their orientations pointing along the connection axis. This has implications on the interpretation and theoretical description of the PMFs along their COM distance reaction coordinate, as discussed in the following. 

\subsection{PMF profiles and counterion-release}

\begin{figure}[h]
\includegraphics[scale=0.65]{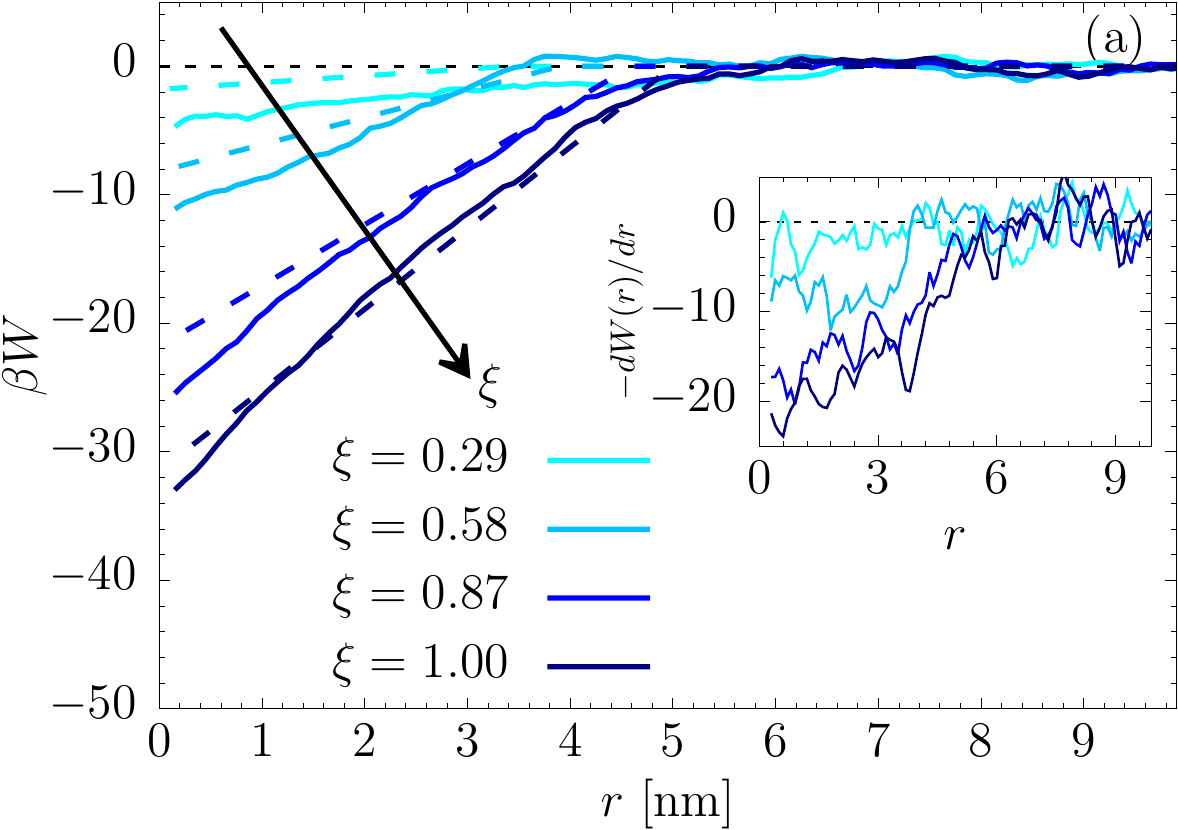}
\includegraphics[scale=0.65]{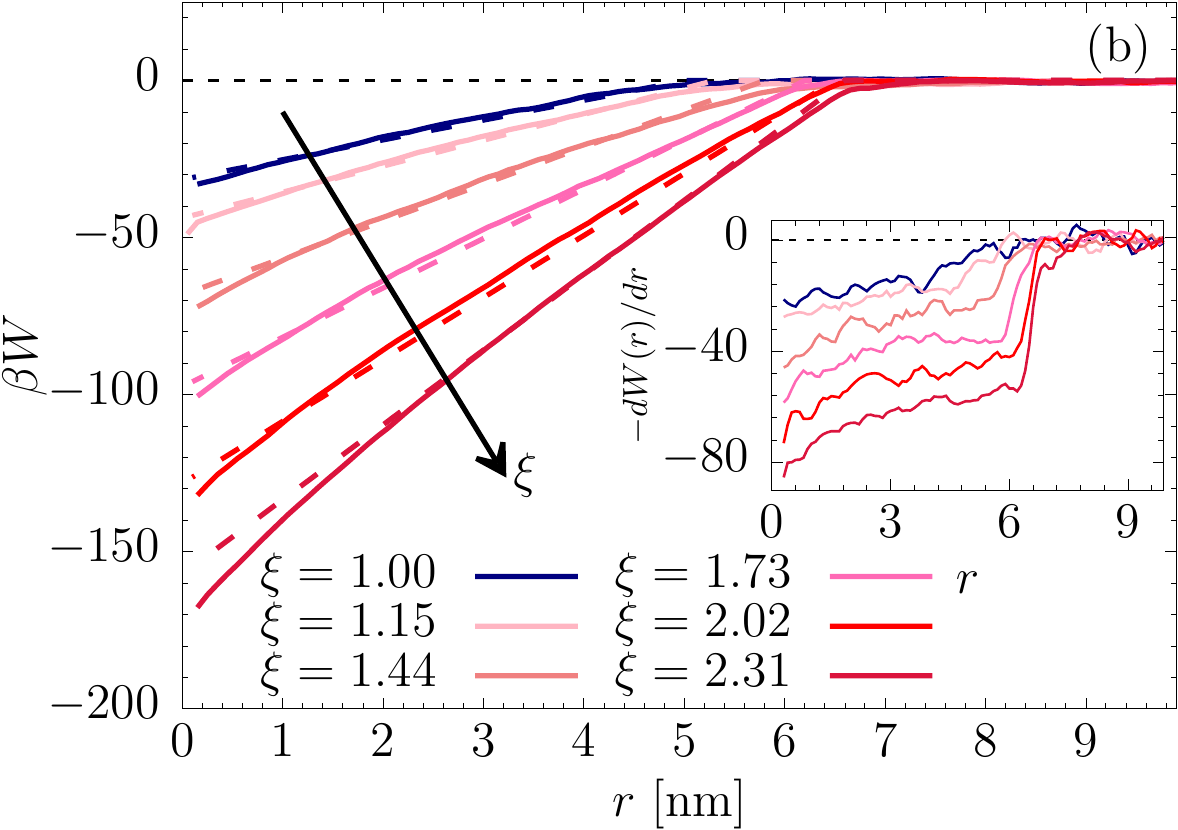}
\caption{
The PMF $W(r)$ in units of $k_{\textrm B}T=\beta^{-1}$ between two oppositely charged PEs as a function of
their COM distance $r$. The simulation results  are represented by solid curves for different values of  the Manning parameter $\xi$ (see legend).
The bold dashed lines are the PMF predictions $W_{\textrm {theo}}(r)$ from \Eq~(\ref{fr_total}). 
The insets show the mean force  $-dW(r)/dr$ in units of $k_{\rm B}T$\,nm$^{-1}$. The thin dashed line is the baseline $W(r) = 0$. The arrows signify the trend for increasing $\xi$.
}
\label{pmf}
\end{figure}

In \Fig~\ref{pmf} we present the PMF, $W(r)$, between the two PE chains with different Manning parameters $\xi$.  Here the salt concentration is again $c_0\simeq20$ mM. Figure~\ref{pmf}(a) shows the PMFs for $\xi\leq 1$ and (b) for $\xi \geq 1$. The three generic stages of the PE--PE association discussed before are reflected also in the behavior of the PMF: When the PE chains are far apart, $r\gtrsim 6.5$\,nm, they do not significantly interact with each other (on the shown scale) regardless of the Manning parameter. Once the PE chains begin to associate at $r^* = 5-7$~nm and stretch at intermediate separations $1-2$\,nm $< r < r^*$, $W(r)$ becomes nearly a linear function of $r$, with the slopes increasing with increasing $\xi$.  Consequently, 
the mean force $f = - dW(r)/dr$, shown in the insets, can be regarded as nearly constant  in that $r$-range, as reported already in previous 
simulations~\cite{Peng2015}. However, some slope in the mean force is clearly visible, especially for the smaller $\xi$-values, so the assumption of a strictly
constant mean force is not generally true.  Strikingly, the mean force $f = - dW(r)/dr$ exhibits a discontinuity at $r^*$ for $\xi\gtrsim 1$, see the inset to \Fig~\ref{pmf}(b), a fact that will be discussed later in more detail.  At smaller separations $r< 1-2$\,nm, the PE chains tend to intertwine into a collapsed globule and by that further increasing the association free energy. In this regime, the corresponding attraction grows even stronger (superlinear) with the distance. The value of $W(r\simeq 0)$ at the associated state represents the free energy of PE--PE complexation. 
Its value increases with the Manning parameter $\xi$, as can be expected from increased Coulomb attraction between both PEs.
The thermodynamics in terms of enthalpy and entropy of the final complex was investigated in detail by Ou and Muthukumar~\cite{Ou2006}.

In the same figures, we plot the PMF predictions of our simple theory given by \Eq~(\ref{fr_total}) by dashed lines. We note again that for each $\xi$, we determine the effective length $L_r $ at the onset of attraction of stretched PEs directly from \Fig~\ref{re}(a) and use it as an input to the theory, which is applicable for $r<L_r$.  Another parameter in our analytical model, the rod distance $r_\perp=0.28$~nm, is fixed by fitting the theory to the linear part of the PMF for $\xi=1$, where PE chains are mostly stretched and the theory is expected to be most reliable. For all other $\xi$-values the model now delivers a prediction.  We see in  \Fig~\ref{pmf}(a)  that for  $\xi \le 1$, the discrepancy between the theory and the simulations grows in relative terms with decreasing $\xi$ at the whole attractive range of $r$. This is expected as the assumption of a parallel rod-like sliding mechanism becomes less accurate for decreasing $\xi$.  In contrast, for $\xi > 1 $ the linear trend of the analytical prediction agrees very well, even quantitatively, with the simulated PMFs for intermediate distances $2.5$~nm $\lesssim r \lesssim 6.5$~nm. Note that in this regime the mean force is relatively constant, as assumed in the theory. At very small distances $r\lesssim 2$\,nm, the PE chains collapse into a globule and therefore the rod model clearly breaks down. This shows that the first steps of the PE association for highly-charged PEs can be very well captured by the analytical model based on two rigid sliding rods together with counterion release entropy. 
Note that the differences between the PMF for $\xi=1$ and the PMFs for $\xi>1$ are solely provided by counterion-release entropy.
The enthalpic Coulomb part (attraction of rods with renormalized charge) becomes less important with increasing $\xi$. 
This is consistent with the findings for the thermodynamics of the complexed state~\cite{Ou2006}.

\begin{figure}
\includegraphics[scale=0.65]{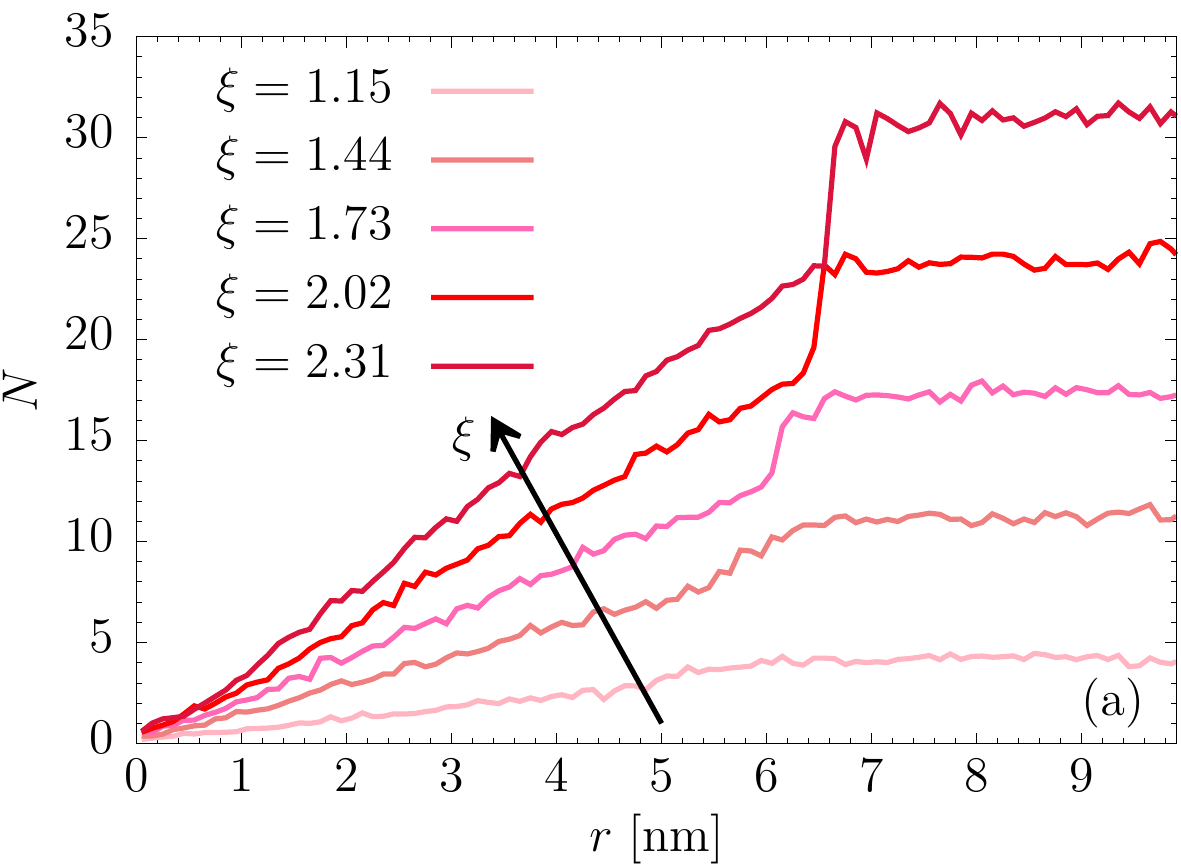}
\includegraphics[scale=0.65]{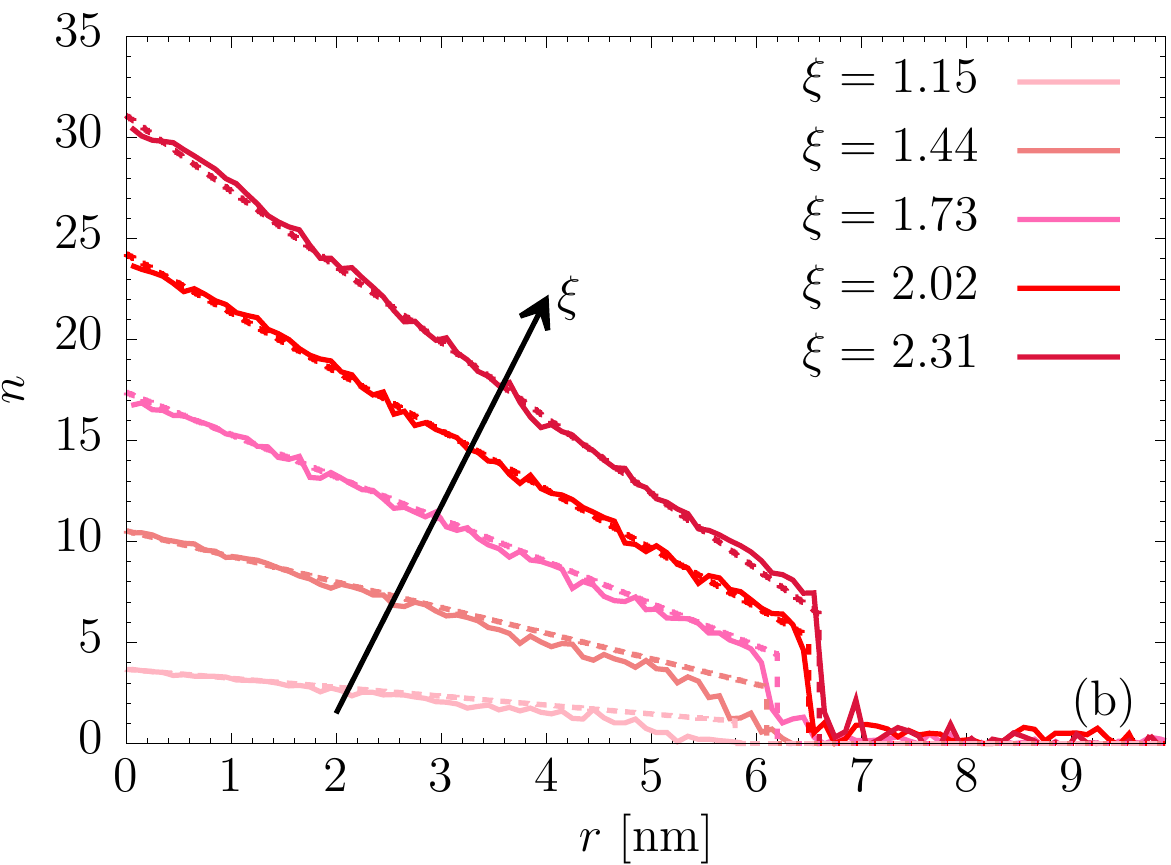}
\caption{(a) The total number of condensed counterions $N$, and  (b) the total number of the released counterions $n$ per PE versus the
PE--PE COM distance $r$ during PE complexation. (The numbers {\it per PE} are half of that.) 
The simulation results are represented by solid curves for different values of the Manning parameter $\xi$ (see legend).
The dashed lines are the number of released counterions predicted via \Eq~(\ref{nnr}). The arrows signify the trend for increasing $\xi$.}
\label{ions}
\end{figure}

As a minor but interesting note we now estimate the relative time period of the association process into the final complex with respect to ordinary 
diffusion. The typical diffusion time over a length $L$ in a simple Rouse picture~\cite{Rouse} would be $\tau_D \simeq N_b \xi_m L^2/k_BT$, where $\xi_m$ is the friction constant of a monomer.  
This has to be compared to a macromolecule associating a length $L$ with a speed $L/\tau_{\rm assoc} = f/(N_b\xi_m)$ under the influence
of a driving force $f$.  Comparing the time scales, we obtain simply $\tau_{\rm D}/\tau_{\rm assoc} \simeq \beta f L$. 
Pluggin in our calculated values of the mean force (insets to Fig.~3) for $L$ on a nanometer scale we see that, for not too small 
Manning-parameters $\xi$, the association for electrostatic- and counterion-release driven PE complexation time can be 
easily 1-2 orders of magnitude smaller than a simple diffusion-dominated association process. These short time scales have
indeed been observed in coarse-grained computer simulations of unrestrained PE complexation~\cite{Muthu:1994, Winkler:2002,Ou2006}.  

In order to further corroborate our theoretical assumptions on the counterion release effect, we now examine the number 
of released ions during the association process. Figure~\ref{ions}(a) presents the number $N(r)$ of condensed counterions per PE (averaged over both chains) for various $\xi$  values resolved in COM distance $r$. In the coil phase, where both PEs are independent of each other, the amount of condensed counterions reaches its maximal value.  As mentioned before, we define the Manning radius $r_0$ for each value of $\xi$ such that the theoretical prediction for $N = N_b z_b (1-1/\xi)$ can be reproduced for an isolated rod.  
At the critical distance $r^*$, the  number of condensed counterions experiences a discontinuous jump $\Delta N$. 
Here, the coil phase goes over to the elongated phase where the end parts of the PEs stick together and by that trigger the release of $\Delta N$ counterions. The latter increases with $\xi$, where in the case of $\xi = 2.31$ more than $3$ counterions are liberated per PE. After the jump, a linear decrease in the number of condensed counterions demonstrates the progressive release of counterions from the overlapping segments of PEs, as predicted by our theory, \Eq~(\ref{nnr}). Finally, all counterions are released in the final complex at $r=0$, where both PEs completely neutralize each other. Here, we can observe that between $\xi=1.15$ and $2.31$ between 2 and 15.5 ions are released per PE into the bulk, respectively. Assuming approximately 5~$k_{\rm B}T$ per ion (see methods after eq.(16)) in the case of $\xi=2.31$, we end up with a total free energy of complexation about 155 $k_{\rm B}T$, in good agreement with 
the PMF data in  \Fig~\ref{pmf}(b). 

\subsection{Discontinuity in PE complexation}

\begin{figure}[ht]
\includegraphics[scale=0.65]{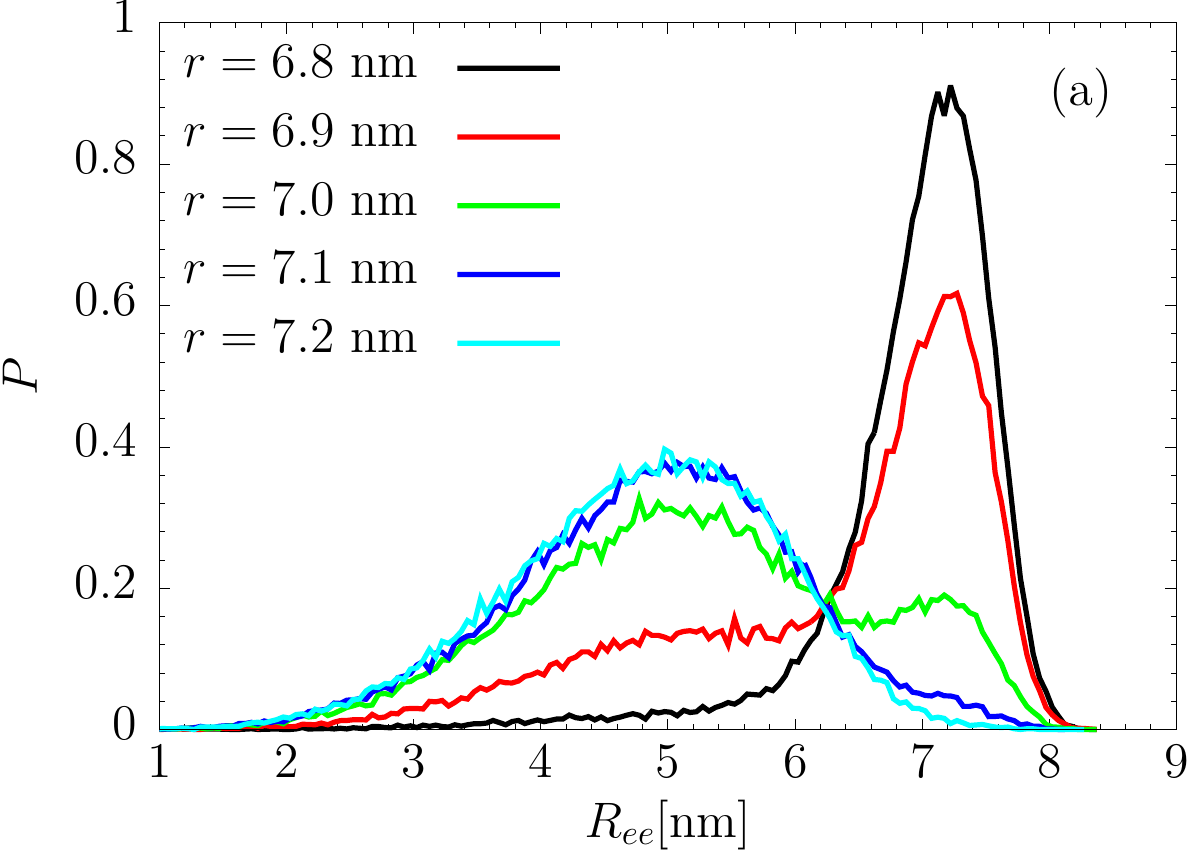}
\includegraphics[scale=0.63]{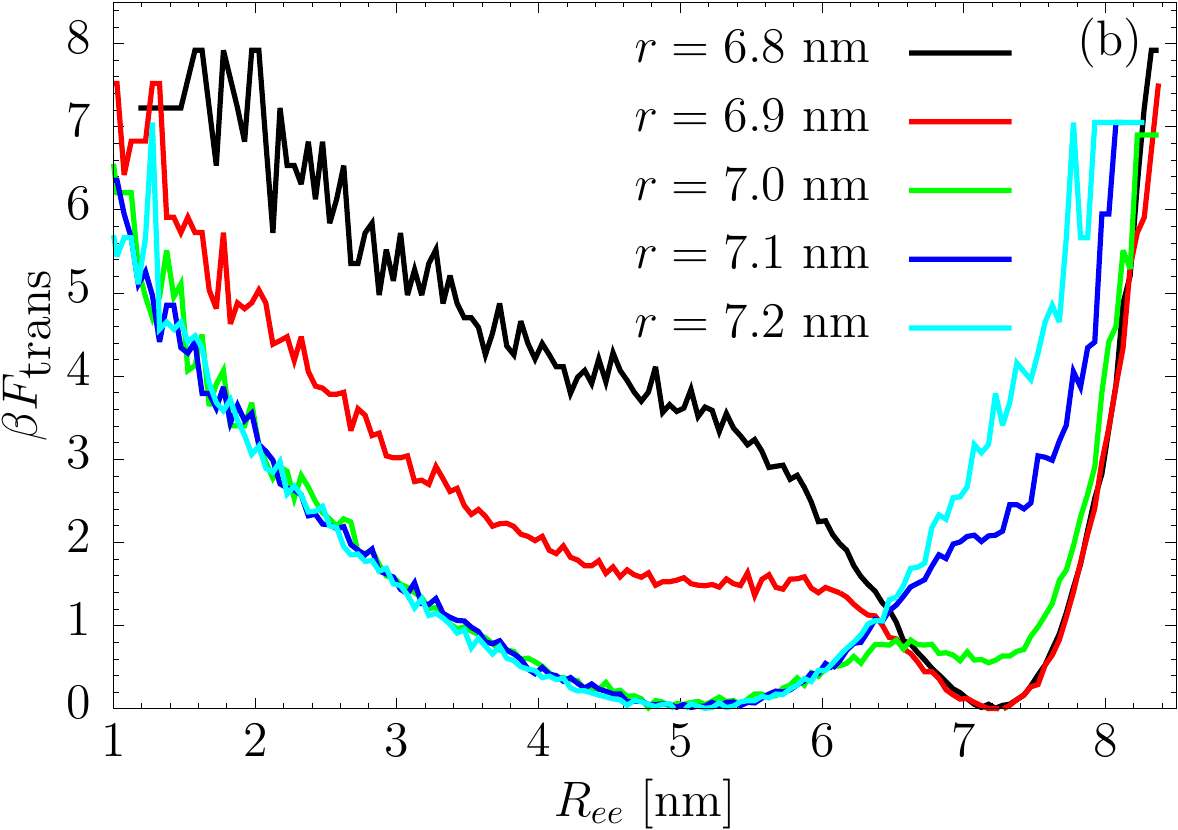}
\caption{(a) Probability distribution $P(R_{ee})$ for different COM distances $r$ close to the critical jump distance $r^*\simeq 6.9-7.0$~nm with the Manning parameter $\xi = 2.31$. 
(b) The corresponding transition free energy $F_{\textrm {trans}} = -k_{\textrm B}T\ln P$ as a function of $R_{ee}$.}
\label{his}
\end{figure}

The discontinuities in the mean force, condensed counterion number $N(r)$, as well as the end-to-end distance $R_{ee}$  clearly suggest that the system undergoes an abrupt change at the critical distance $r^*$ of the approaching PE chains.  In an attempt to characterize this transition in more detail, we have calculated the probability distribution $P(R_{ee})$ of the PE end-to-end distance for various fixed (harmonically constrained)  COM distances $r$ in the range from $6.8$\,nm to $7.2$\,nm for the case of $\xi=2.31$, thereby crossing its critical distance $r^*$, shown in Fig.~\ref{his}(a). For the two largest as well as for the smallest distance $r$, we find single-peaked distributions, corresponding  to the well defined  single states,  coil and stretched PE chains, respectively. At around $r^* = 6.9-7.0$\,nm, however,  a bimodal distribution between the two states appears, indicating a structural coexistence, separated by unlikely states. The transition free energy profiles $F_{\textrm {trans}}(R_{ee}) = -\kB T\ln\,P(R_{ee})$ are plotted in \Fig~\ref{his}(b). In the bimodal states, the potential barrier separating the two states has a height of the order of $\Delta F_{\textrm {trans}} \sim \kB T$. Note also that the peak position of the extended state is located at values of about $L_r \simeq 7.2$~nm, somewhat larger than $r^*$. The reason is that the stable state apparently needs some finite overlap, i.e., a critical attraction to warrant a stable state.

\begin{figure}[h]
\includegraphics[scale=0.7]{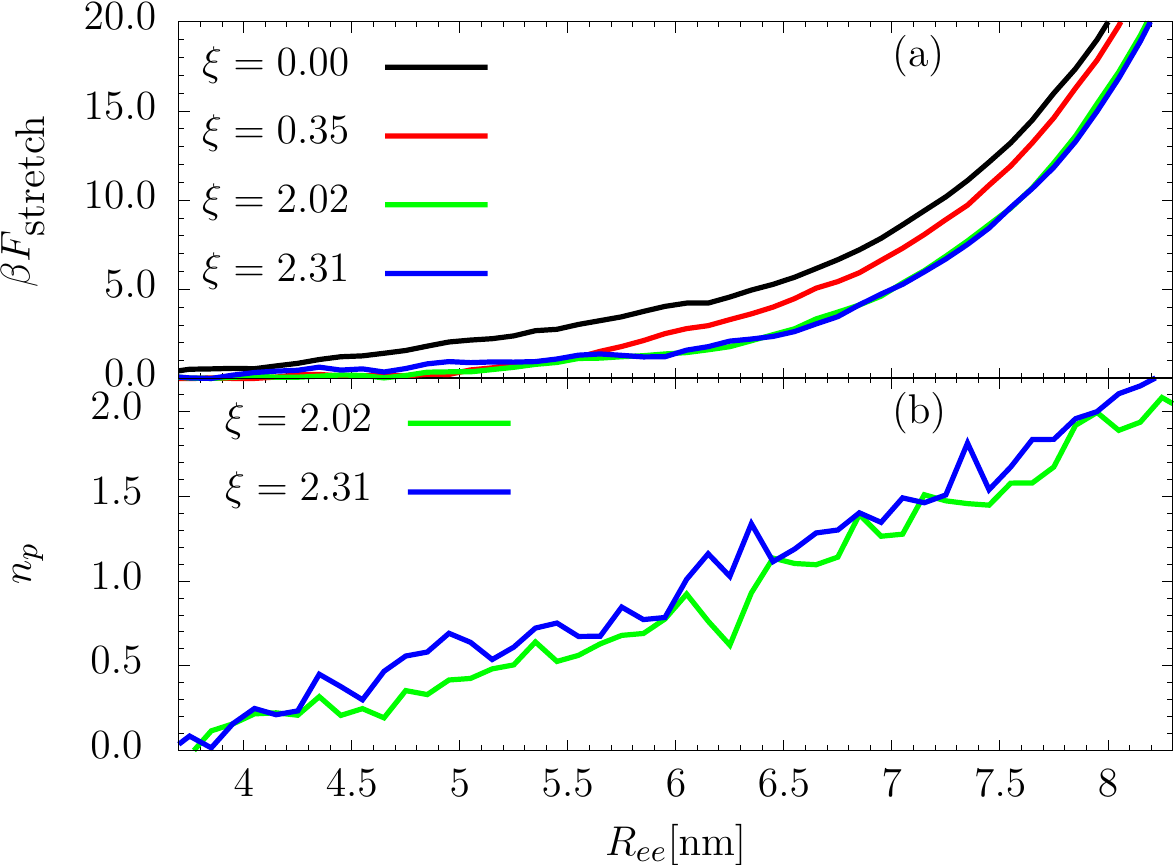}
\caption{ (a) The free energy of stretching a single PE is plotted as a function of $R_{ee}$ as obtained from pulling simulations.
(b) The number of the released counterions $n_p$ during stretching a single PE is plotted as a function of its end-to-end distance $R_{ee}$. }
\label{pull}
\end{figure}

So, what is the reason for this bimodal distribution? Consider first the PE chains in the coil state, i.e., when $r>r^*$. In a rare fluctuation, 
the chains stretch out, accompanied by a significant loss in conformational entropy, and may achieve their handshake by overlapping with 
one or a few more monomers at the ends. For $\xi>1$, we have shown that this 'first touch' will be accompanied by a significant 
release of counterions, contributing a large favorable entropy of about 5~$k_{\rm B}T$ per released ion.  Apparently, this gain in counterion entropy is large enough to compensate for the loss in the conformational entropy of PE chains such that a coexistence can be established in this restrained equilibrium. Quantitatively, we estimate the stretching entropy of a single PE from our steered Langevin simulations. Results are shown in Fig.~\ref{pull}(a) for various $\xi$ values, including the neutral reference $\xi=0$. For $\xi>1$, the PE stretching is a bit easier than for a neutral polymer owing to the internal electrostatic repulsion. However, for example for $\xi=2.31$, we measure an appreciable entropy loss of about 10~$k_{\rm B}T$ per PE chain for the extension $L_r\simeq 7.5$~nm. The loss in the chain conformation entropy can be easily compensated by the handshake and forthcoming release of counterions. For this particular case, thus around two times ten, i..e, 20~$k_{\rm B}T$ in total, the release of about four counterions, would be needed to establish the coexistence between associated and free states. Glancing back at \Fig~\ref{ions}, however, we see that for $\xi=2.31$ in total seven ions are released at the handshake. The apparent discrepancy can be reconciled by investigating the number of released counterions during stretching,  see Fig.~\ref{pull}(b). In the stretched case, $1.5$ ions per chain (thus, three in total) are released on average for $L_r\simeq 7.5$~nm for $\xi =2.31$. In other words, only four counterions are released in the actual handshake, in agreement with the needed compensation stretching penalty mentioned above. The subtle structural effects on a single-ion level are due to the internal structure and flexibility of the PEs and are beyond the scope of the traditional Manning theory, but, as we show here, are important for a quantitative interpretation of the transient states in complexation. The free energy barrier between the coil and extended state must then be clearly attributed to the entropy of intermediate stretching.

\section{Summary and concluding remarks}

In summary, we have studied PE structure variations and the resultant PMF profiles along the PE--PE center-of-mass reaction coordinate for PE pair complexation with a focus on intermediate association ranges for various PE charge densities.
For charge densities above the condensation threshold, we observed and analyzed in detail a (fast) sliding-rod-like process 
preceding the PE complexation. We introduced an abstract model leading to an analytical expression for the PMF. The latter predicts 
a PMF virtually linear in center-of-mass distance below the onset of complexation until collapsing into the final complex, 
in good agreement with the computer simulations. Furthermore, a detailed inspection of the mean force profile  uncovered a 
discontinuity at the onset of complex formation, which is also embodied as a jump in a number of simulation measures.  
We demonstrated that  the discontinuity can be attributed to the presence of a free-energy barrier stemming from cooperative
counterion-release effects and single PE stretching entropy.

Because of the drastic changes in the topology of polymer chains, we suspect that the metastable states that emerge preceding to PE 
complexation may be relevant for realistic biological systems.  For example, polycations for gene delivery are designed to complex with 
nucleic acids such that the gene transfer vectors are able to overcome the intracellular barriers such as the plasma membrane, the endosome 
and the nuclear membrane. Successful gene delivery depends upon the ability of the vector to adopt different metastable structures or 
possess certain properties at different intracellular environments~\cite{Kissel, JihanZhou2013, Ziebarth2009}. Here, one could suspect
that transiently stretched or coiled states could provide some function, possibly some disorder-order based signaling, in a specific environment. 
In that respect it  is also interesting to note that  transient disorder-to-order transitions, coupled with the adoption of different structures with different partners, 
are a common feature of intrinsically disordered proteins (IDPs) whose capacity for binding diversity plays important roles in both protein-protein interaction networks and likely also in gene regulation networks. 

We finally note that the discontinuity observed in the PE-PE association process may also emerge during the complexation of PEs and globular proteins as indicated in recent coarse-grained computer simulations of more~\cite{ShunYu2015} or less~\cite{CemilYigit2015} resolved protein models 
with heterogeneous charge distributions.  In these cases, counterion-release effects after PE binding to protein surface charge patches 
have also been identified as the main non-specific interaction  responsible for the PE-protein complexation.  

\section{Acknowledgments}
X. X. acknowledges funding from the Chinese Scholarship Council (CSC). Research at UCR is financially supported by the U. S. National Science Foundation (NSF-CBET-0852353). M. K. and J.D. acknowledge funding by the ERC (European Research Council) Consolidator Grant with project number 646659--NANOREACTOR. 

\newpage

\bibliography{PEPE}

\end{document}